\documentclass[aip,apl,reprint]{revtex4-1} 
\usepackage{graphicx}
\usepackage{dcolumn}
\usepackage{bm}
\usepackage{amsmath, amssymb}

\usepackage{amsfonts, amsmath, bm, graphicx}

\def\vec#1{{\bm{#1}}}
\def\mat#1{{\hat{\vec{#1}}}}
\def\H{{\mathcal{H}}}

\begin{document}

\title{Excitation of  whispering gallery magnons in a magnetic vortex}

\author{K. Schultheiss}
\affiliation{Helmholtz-Zentrum Dresden--Rossendorf, Institute of Ion Beam Physics and Materials Research, Bautzner Landstra\ss e 400, 01328 Dresden, Germany}

\author{R. Verba}
\affiliation{Institute of Magnetism, National Academy of Sciences of Ukraine, Kyiv 03680, Ukraine}

\author{F. Wehrmann}
\affiliation{Helmholtz-Zentrum Dresden--Rossendorf, Institute of Ion Beam Physics and Materials Research, Bautzner Landstra\ss e 400, 01328 Dresden, Germany}

\author{K. Wagner}
\affiliation{Helmholtz-Zentrum Dresden--Rossendorf, Institute of Ion Beam Physics and Materials Research, Bautzner Landstra\ss e 400, 01328 Dresden, Germany}
\affiliation{Technische Universit\"at Dresden, 01062 Dresden, Germany}

\author{L. K\"orber}
\affiliation{Helmholtz-Zentrum Dresden--Rossendorf, Institute of Ion Beam Physics and Materials Research, Bautzner Landstra\ss e 400, 01328 Dresden, Germany}
\affiliation{Technische Universit\"at Dresden, 01062 Dresden, Germany}

\author{T. Hula}
\affiliation{Helmholtz-Zentrum Dresden--Rossendorf, Institute of Ion Beam Physics and Materials Research, Bautzner Landstra\ss e 400, 01328 Dresden, Germany}
\affiliation{Wests\"achsische Hochschule Zwickau, 08056 Zwickau, Germany}

\author{T. Hache}
\affiliation{Helmholtz-Zentrum Dresden--Rossendorf, Institute of Ion Beam Physics and Materials Research, Bautzner Landstra\ss e 400, 01328 Dresden, Germany}
\affiliation{Technische Universit\"at Chemnitz, 09111 Chemnitz, Germany}

\author{A. Kakay}
\affiliation{Helmholtz-Zentrum Dresden--Rossendorf, Institute of Ion Beam Physics and Materials Research, Bautzner Landstra\ss e 400, 01328 Dresden, Germany}


\author{A.A. Awad}
\affiliation{Department of Physics, University of Gothenburg, 412 96 Gothenburg, Sweden}

\author{V. Tiberkevich}
\affiliation{Department of Physics, Oakland University, Rochester, MI 48309, USA}

\author{A.N. Slavin}
\affiliation{Department of Physics, Oakland University, Rochester, MI 48309, USA}

\author{J. Fassbender}
\affiliation{Helmholtz-Zentrum Dresden--Rossendorf, Institute of Ion Beam Physics and Materials Research, Bautzner Landstra\ss e 400, 01328 Dresden, Germany}
\affiliation{Technische Universit\"at Dresden, 01062 Dresden, Germany}

\author{H. Schultheiss}
\affiliation{Helmholtz-Zentrum Dresden--Rossendorf, Institute of Ion Beam Physics and Materials Research, Bautzner Landstra\ss e 400, 01328 Dresden, Germany}
\affiliation{Technische Universit\"at Dresden, 01062 Dresden, Germany}

\date{\today}

\maketitle
{\bf One of the most fascinating topics in current quantum physics are hybridised systems, in which different quantum resonators are strongly coupled. Prominent examples are circular resonators with high quality factors that allow the coupling of optical whispering gallery modes \cite{Mie,Debye,Cai,Oraevsky,Vahala} to microwave cavities \cite{TabuchiPRL} or magnon resonances in optomagnonics \cite{Haigh2016,Kusminskiy, Osada}. 
Whispering gallery modes play a special role in this endeavour because of their high quality factor and strong localisation, which ultimately increases the overlap of the wavefunctions of quantum particles in hybridised systems. The hybridisation with magnons, the collective quantum excitations of the electron spins in a magnetically ordered material, is of particular interest because magnons can take over two functionalities: due to their collective nature they are robust and can serve as a quantum memory\cite{Zhang2015} and, moreover, they can act as a wavelength converter between microwave and THz photons\cite{Kusminskiy}. However, the observation of whispering gallery magnons has not yet been achieved due to the lack of efficient excitation schemes for magnons with large wave vectors in a circular geometry. To tackle this problem, we studied non-linear 3-magnon scattering\cite{suhl1957, ordonez2009, Schultheiss2009, Camley2014} as a means to generate whispering gallery magnons. This Letter discusses the basics of this non-linear mechanism in a confined, circular geometry from experimental and theoretical point of view.}

\begin{figure}
\begin{center}
\scalebox{1}{\includegraphics[width=7.5cm]{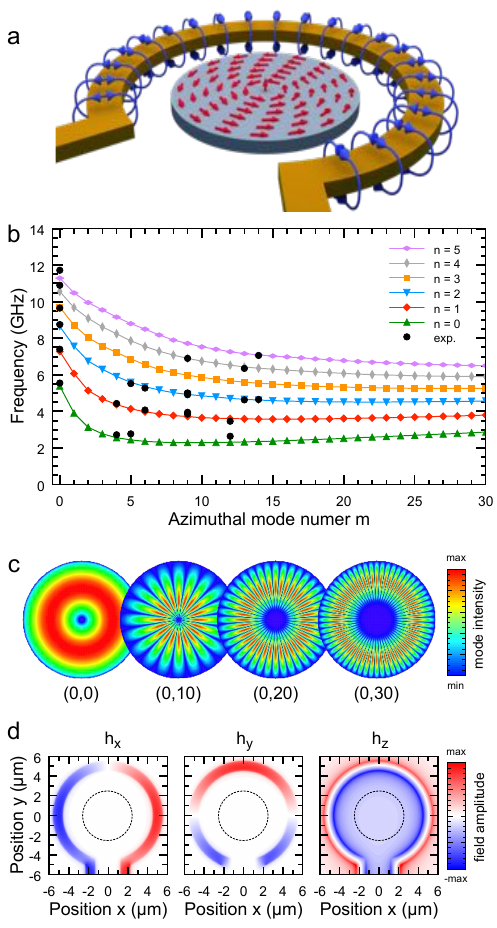}}
\caption{\label{fig1}\textbf{Sample layout and characteristics of magnon whispering gallery modes.} {\bf a}, A Ni$_{81}$Fe$_{19}$ disc with 50\,nm thickness and 5.1\,$\mu$m diameter is patterned inside an $\Omega$-shaped antenna. Red arrows depict the magnetisation configuration of the magnetic vortex structure and the blue lines represent the dynamic magnetic field generated by the loop shaped microwave antenna. {\bf b}, Analytical calculation of the magnon mode frequencies as a function of the radial and azimuthal mode numbers $n$, $m$ (see methods). Black dots show experimental results. {\bf c}, Four exemplary mode profiles. The larger $m$, the more pronounced the character of the whispering gallery magnons is revealed. {\bf d},  COMSOL simulation of the $x-$, $y-$, and $z$-component of exciting magnetic field {\bf h} generated by the $\Omega$-shaped antenna. The dashed circle indicates the size and position of the disc. }
\end{center}
\end{figure}

Whispering gallery magnons can only live in systems with rotational symmetry. This not only applies to the geometry of the magnetic element but also to the magnetisation texture therein. For that reason, we study  a Ni$_{81}$Fe$_{19}$ disc that inherently exhibits a magnetic vortex structure\cite{Shinjo,Novosad2002,Guslienko2004,Novosad2005,Guslienko2008}. The arrows  in Fig.~\ref{fig1}a depict the generic features of such a vortex state: the magnetic moments curl in plane along circular lines around the vortex core, a nanoscopic region in the center of the disc where the magnetisation tilts out of plane. According to this rotational symmetry, the magnon eigenmodes in a vortex are characterised by mode numbers ($n,m$), with $n = 0, 1, 2, ...$  counting the number of nodes across the disc radius and $m = 0, \pm1,  \pm2, ...$  counting the number of nodes in azimuthal direction over half the disc \cite{Kalinikos1986, Buess2004}.

Other than commonly known waves, like sound, water or electromagnetic waves, magnons exhibit a strongly anisotropic dispersion relation in in-plane magnetised thin films\cite{Kalinikos1986}. In a vortex, this results in increasing (decreasing) mode energies for increasing $n$ ($m$) as shown by the analytic calculations in Fig.~\ref{fig1}b. The four exemplary intensity profiles for the eigenmodes $(0,0)$, $(0,10)$, $(0,20)$, and $(0,30)$, that are shown in Fig.~\ref{fig1}c, reveal the character of whispering gallery magnons: the larger $m$, the more the magnon intensity is pushed toward the perimeter of the disc which can be understood intuitively by the reduction of exchange energy: Leaving an extended area around the vortex core with zero amplitude avoids a strong tilt of neighbouring spins close to the vortex core and, therefore, reduces the total energy.

Even though magnon spectra in magnetic vortices have been intensively studied in the past \cite{Buess2004, Zivieri2005, Guslienko2008, Awad2010}, magnons with large azimuthal wave vectors have not yet been measured experimentally and were only observed in micromagnetic simulations\cite{Taurel2016}. The challenge to  generate such magnons and, thereby, to reach out to whispering gallery magnons is finding an efficient excitation mechanism in a micron-sized vortex. Here, we tackle this problem via nonlinear 3-magnon scattering. In this process, one magnon splits in two new magnons under conservation of energy and momentum. The rotational symmetry of the vortex texture implies specific selection rules for the scattering process which we will describe in context with the experimental data.

In order to selectively drive magnetisation dynamics, we apply microwave currents to an $\Omega$-shaped antenna that encloses the vortex. Inside the $\Omega$ loop, a  spatially uniform magnetic field is generated that is oriented perpendicularly to the disc  as  shown in Fig.~\ref{fig1}d. The rotational symmetry of this magnetic field prohibits direct coupling to magnons with $m\neq0$. However, because of the small diameter of the antenna, strong magnetic fields can be generated so that these magnons can be indirectly driven in the nonlinear regime via multi-magnon scattering processes.

\begin{figure*}
\begin{center}
\scalebox{1}{\includegraphics[width=17cm]{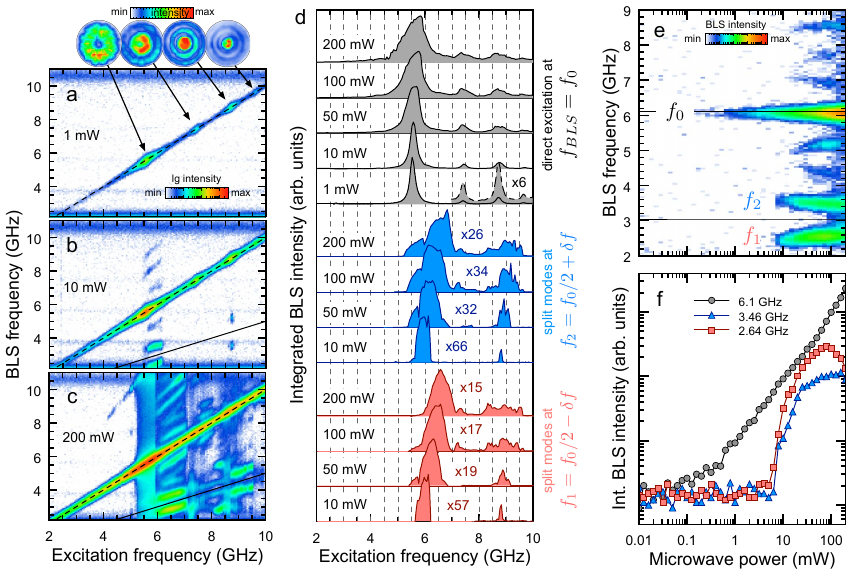}}
\caption{\label{fig2}\textbf{Magnon excitation spectra for different applied microwave powers.}~{\bf a}-{\bf c}, BLS spectra for microwave frequencies $f_{0}$ between 2 and 10\,GHz and excitation powers of 1, 10, and 200\,mW, respectively. The diagonal dashed lines indicate the directly excited magnetic oscillations at $f_\mathrm{BLS}=f_\mathrm{0}$. At 10 and 200\,mW, off-diagonal signals associated with multi-magnon scattering processes are detected. {\bf d}, Black data show the BLS intensity integrated in 800-MHz wide windows around the direct excitation for 1, 10, 50, 100, and 20\,mW (bottom to top). At 1\,mW, the intensity integrated for excitation frequencies between 7 and 10\,GHz was multiplied by a factor of six to better see the resonances of the higher order radial modes. Blue (red) data  show the intensities of the split modes integrated in 1.4-GHz wide windows around  $f_\mathrm{1}=f_\mathrm{0}/2+ \delta f$ ($f_\mathrm{2}=f_\mathrm{0}/2- \delta f$) with $\delta f = 800\,\mathrm{MHz}$. {\bf e}, Power dependence of the BLS spectra excitated at $f_\mathrm{0}=6.1$\,GHz. {\bf f}, BLS intensity integrated in 800-MHz wide frequency windows around the BLS frequencies $f_\mathrm{0}=6.1$\,GHz, $f_\mathrm{1}=2.64$\,GHz, and $f_\mathrm{2}=3.46$\,GHz as a function of the microwave power. In the double-logarithmic plot, the direct excitation at $6.1$\,GHz follows a linear trend, whereas the split modes at 3.46 and 2.64\,GHz show a clear threshold behaviour.
}
\end{center}
\end{figure*}

We track these nonlinear processes by measuring magnon spectra as a function of the applied microwave frequency using Brillouin light scattering (BLS) microscopy\cite{Sebastian2015}. We would like to emphasise that even though the system is driven with one microwave frequency at a time, the BLS technique allows us to detect the dynamic magnetic response in a broad frequency range. In Fig.~\ref{fig2}a-c  we plot the measured BLS spectra between 2 and 11\,GHz ($y$-axis) for each excitation frequency ($x$-axis) at microwave powers of 1, 10, and 200\,mW. The magnon intensity is encoded using the same logarithmic scale shown as an inset in Fig.~\ref{fig2}a.

At the lowest microwave power of 1\,mW (Fig.~\ref{fig2}a), magnons are driven in the linear regime, which is corroborated by the fact that magnons are only observed at the BLS frequency that matches the applied microwave frequency $f_\mathrm{BLS}=f_\mathrm{0}$. Hence, the measured intensities strictly follow the diagonal, dashed line. Four distinct resonances emerge at 5.55, 7.40, 8.75, and 9.65\,GHz, which we identify as the well known first four radial modes\cite{Buess2004, Vogt2011} by spatially-resolved BLS microscopy (insets in Fig.~\ref{fig2}a).

For a power of 10\,mW (Fig.~\ref{fig2}b) the excitation field is already strong enough to drive magnons in the nonlinear regime. Hence, we observe strong off-diagonal signals that appear at BLS frequencies symmetrically spaced around half the excitation frequency $f_\mathrm{0}/2$ (straight line with slope 0.5). These satellite peaks are the result of 3-magnon splitting processes. In order to conserve energy the initial magnon with frequency $f_{0}$ splits in two magnons with frequencies $f_{1}=f_{0}/2-\delta f$ and $f_{2}=f_{0}/2+\delta f$. Moreover, the rotational symmetry of the vortex requires conservation of the momentum component in azimuthal direction. For an initial magnon with $m=0$ this implies that the split modes have azimuthal mode numbers with the same modulus but opposite sign: $m_{1}=-m_{2}$. Our analytic calculations further show that the split modes may not share the same radial index, {\it i.e.}, $n_{1} \neq n_{2}$ (see methods). All three selection rules  drastically restrict the possible scattering channels within the discrete eigenmode spectrum of the vortex (see Fig.~\ref{fig1}b).

At the maximum applied microwave power of 200\,mW, the number of off-diagonal signals increases further (Fig.~\ref{fig2}c). Especially, for excitation frequencies between $6$ and $7$\,GHz, we do not just measure two satellite peaks with frequencies $f_{1}$ and $f_{2}$ but a total number of ten additional modes. Their presence is attributed to avalanche processes of higher-order multi-magnon scattering. Their frequencies are given by combinations of the three initial magnons, e.g., $2 f_{1}$, $2 f_{2}$, $f_{0}+f_{1}$. Furthermore, the significant line broadening (additional noise) of the directly excited mode and the split modes in the frequency range between $5.3$ and $5.9$\,GHz can be attributed to 4-magnon scattering\cite{Schultheiss2012}. However, this article solely focusses on the study of the initial 3-magnon scattering process which clearly dominate in intensity. 

To better illustrate the power dependence of the observed modes, we plot the BLS intensity integrated over different frequency windows as a function of the excitation frequency in Fig.~\ref{fig2}d. The black data resembles the BLS intensity of the direct excitation. With increasing power the initially sharp resonances become broader and show the characteristic nonlinear foldover to higher frequencies \cite{Suhl, Janantha}. The red and blue data in Fig.~\ref{fig2}d show the intensities of the split modes below (red data) and above $f_{0}/2$ (blue data), which overall broaden in range and shift to higher frequencies with increasing power. 

To further elucidate the threshold character of the 3-magnon splitting, we plot a more detailed power dependence of the magnon intensities in Fig.~\ref{fig2}e for  $f_\mathrm{0}=6.1$\,GHz. While the mode at 6.1\,GHz can be observed over a large power range it is evident that the split modes $f_{1}$ and $f_{2}$ appear only above a certain threshold power. Furthermore, we observe a pronounced frequency shift of these two split modes with increasing microwave power. For a quantitative comparison, we integrate the BLS intensity in narrow frequency windows around the directly and indirectly excited modes, respectively (Fig.~\ref{fig2}f). The double-logarithmic scale reveals the linear growth of the direct excitation at $6.1$\,GHz starting at 0.1\,mW. However, the intensities of the satellite peaks around $f_\mathrm{1}=2.65$\,GHz and $f_\mathrm{2}=3.48$\,GHz abruptly increase above 10\,mW which demonstrates the threshold character of the splitting process. 

\begin{figure*}
\begin{center}
\scalebox{1}{\includegraphics[width=15cm, clip]{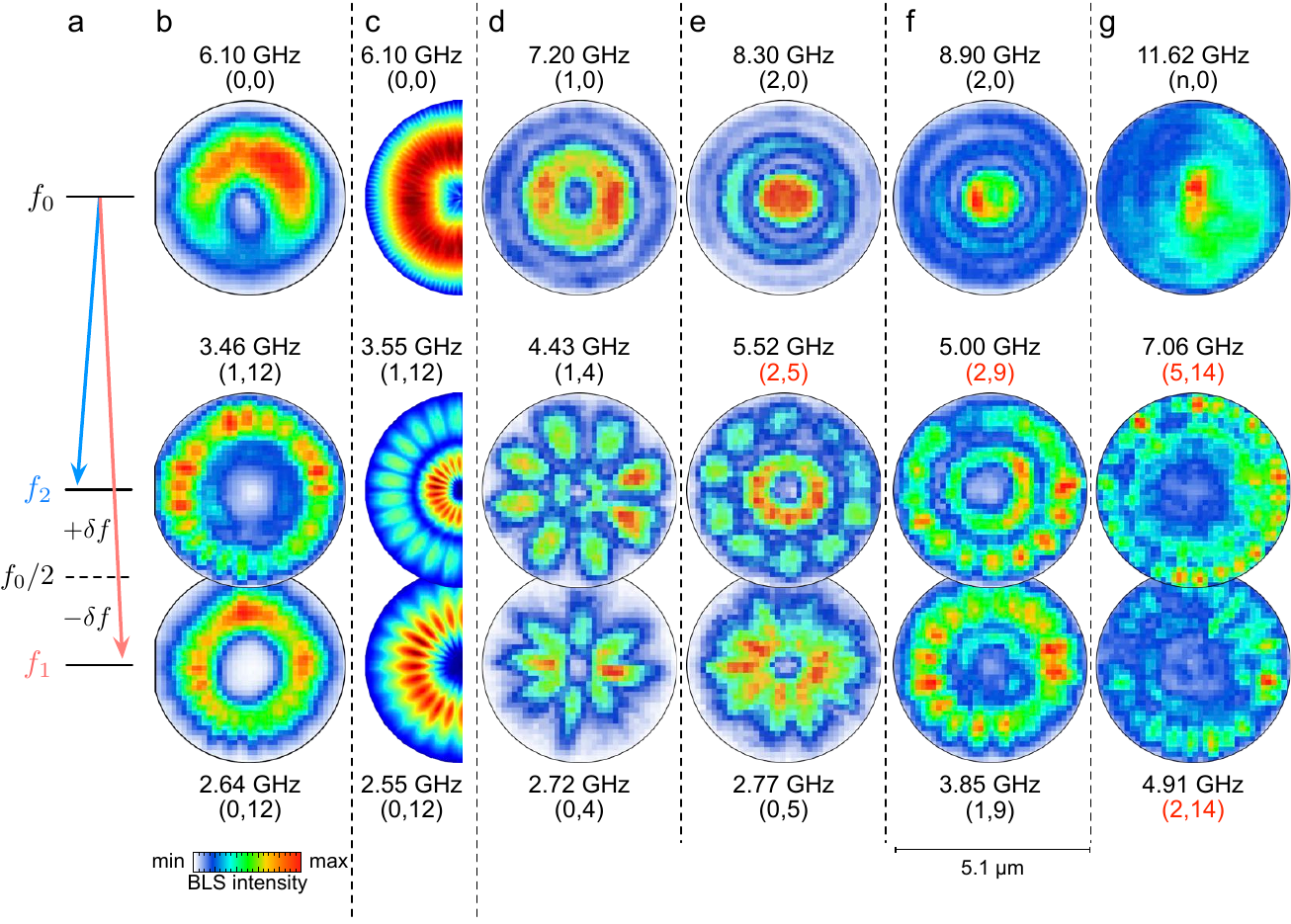}}
\caption{\label{fig3}\textbf{Two-dimensional mode profiles under 3-magnon splitting.} {\bf a}, Energy levels for 3-magnon splitting. {\bf b},{\bf d}-{\bf g}, Spatial intensity distributions of the direct excitation (upper line) as well as the split modes (middle and bottom line) for  various excitation frequencies. {\bf c}, Micromagnetic simulation of the 3-magnon splitting for the excitation frequency of 6.1\,GHz.}
\end{center}
\end{figure*}

In order to reveal the spatial structure of the eigenmodes that are generated via 3-magnon scattering, we simultaneously mapped the profiles of the directly excited mode and the split modes  (Fig.~\ref{fig3}b,d-f). Additionally, we compare the experimental results for the mode with highest intensity at 6.1\,GHz with micromagnetic simulations (Fig.~\ref{fig3}c). The first thing to realise is that all of the split modes show a clear azimuthal character and confirm the analytical calculations and the selection rules imposed by the rotational symmetry: pure radial modes with ($n,0$) split in modes with $m_{1}=-m_{2}$ and $n_{1}\neq n_{2}$. As far as possible, we label the modes according to their radial and azimuthal mode numbers ($n,m$). We resolve azimuthal mode numbers up to 14, to our knowledge the first time to observe vortex modes with such high $m$. For higher $n$, an unambiguous identification of the modes was not possible due to limited spatial resolution. However, the radial mode number can still be retrieved by comparing the measured frequencies to the analytic calculations in Fig.~\ref{fig1}b. We counted the azimuthal mode numbers and plotted the measured frequencies as black dots in the calculated spectrum. From this comparison we then determined the radial mode numbers (red labels in Fig.~\ref{fig3}).
The micromagnetic simulation for excitation at 6.1\,GHz reveals the splitting into magnons with the same mode numbers as in the experiment, however, with slightly different frequencies of the split modes. Reasons for this frequency shift may be attributed to variations in the strength and symmetry of the exciting magnetic field or the material parameters.

Note that we only measure stationary mode profiles which implies that, essentially, all split modes are a superposition of modes counter propagating in azimuthal direction. Therefore, we conclude that the two splitting processes $(n_{0}, 0) \rightarrow (n_{1}, m) + (n_{2}, -m) $ and $(n_{0}, 0) \rightarrow (n_{1}, -m) + (n_{2}, m)$ occur with equal probability. 

It is remarkable, that the higher $m$ for a given $n$, the stronger the mode is localised at the outer circumference of the disc, resembling intensity distributions of optical whispering gallery modes\cite{Yang}. The most beautiful example in our dataset is the intensity distributions of the split mode $(0,12)$ at the excitation frequency 6.1\,GHz shown in Fig.~\ref{fig3}b, which exhibits a distinct hole in its center.

In summary, we shed light on the nonlinear conversion of magnons in a confined system with rotational symmetry by analysing their spectral and spatial characteristics. We showed how this mechanism can be utilised to generate magnons with unprecedented high azimuthal wave vectors and localisation at the discs perimeter, which resembles the character of whispering gallery modes. The underlying 3-magnon scattering processes show a high degree of tunability regarding the frequency and  spatial distribution of the split modes. We believe that this advanced control of the generation of whispering gallery magnons is a missing key towards the realisation of an efficient hybridisation of magnons and other quantum particles as found in circular optical cavities and mechanical quantum resonators.

\section*{Methods}
{\bf Sample Preparation.}  Using electron-beam lithography and conventional lift-off techniques, we patterned magnetic discs with diameters of 5.1\,$\mu$m from a  Ti(2)/Ni$_{81}$Fe$_{19}$(50)/Ti(5) film, that was deposited via electron-beam evaporation on a SiO$_{2}$ substrate. All thicknesses are given in nanometer. In a second  step, we designed  $\Omega$-shaped antennas with an inner and outer diameter of 9 and 11\,$\mu$m, respectively, which separately enclose each individual disc. The antennas are patterned from a Ti(2)/Au(200) layer, also employing e-beam lithography, e-beam evaporation and lift-off techniques, and can be connected to a microwave source via picoprobes.

{\bf BLS microscopy.} All measurements were performed at room temperature. The spin-wave intensity is locally recorded by means of BLS microscopy. This method is based on the inelastic scattering of light and spin waves. Light from a continuous wave, single-frequency 532-nm solid-state laser is focused on the sample surface using a high numerical aperture microscope lens giving a spatial resolution of 250\,nm. The laser power on the sample surface is typically about 1\,mW. The frequency shift of the inelastically scattered light is analysed using a six-pass Fabry-Perot interferometer TFP-2 (JRS Scientific Instruments). To record two-dimensional maps of the spin-wave intensity distribution, the sample is moved via a high-precision translation stage (10-nm step size, Newport). The sample position is continuously monitored with a CCD camera using the same microscope lens. A home-built active stabilisation algorithm based on picture recognition allows for controlling the sample position with respect to the laser focus with a precision better than 20\,nm.

\textbf{Theory.} Here, we briefly describe the derivation of the selection rules for 3-magnon splitting in a vortex-state disc; a detailed and general consideration of this problem will be published elsewhere. We use the standard Hamiltonian formalism for nonlinear magnon interaction (see detailed description, e.g. in Refs.~\onlinecite{Lvov_Book, Krivosik_2010, Livesey_2016}). Neg\-lecting effects of the vortex core (since it results in effects about $R_\mathrm{disc}/R_\mathrm{core} \sim 250$ times smaller than the out-of-core area),  we introduce the canonical variable $a(\vec\rho, t)$ as $-(iM_\rho + M_z) \approx \sqrt{2} M_s a \left(1 - aa^*/4 \right)$, $M_\phi = M_s(1 - aa^*)$, where the polar coordinate system $\rho, \phi, z$ is used \cite{Galkin_2006},  $\vec\rho$ is the in-plane radius vector and $M_s$ is the saturation magnetisation. The canonical variable is expanded into a series of linear magnon modes $a(\vec\rho, t) = \sum_{n, m} a_{nm}(t) g_{nm}(\rho) e^{im\phi}$, where $n = 0,1,2,...$ and $m = 0, \pm1, \pm2, ...$ are the radial and azimuthal mode numbers, $a_{nm}$ is the mode amplitude and the functions $g_{nm}(\rho)$ describe the modes profile in  radial direction.

The Hamiltonian $\H = \gamma W/(M_sV)$ ($W$ is the total magnetic energy and $V$ is the disc volume) in the studied case consists of exchange and dipolar contributions, 
\begin{equation}\label{e:H}
\begin{split}
    \H_\mathrm{ex} & = - \frac{\gamma \mu_0 \lambda_{ex}^2}{2M_s \pi R^2} \int \sum\limits_{\alpha = x,y,z} (\nabla M_\alpha)^2 d\vec\rho \,, \\
   \H_\mathrm{dip} & = \frac{\gamma \mu_0}{2M_s \pi R^2} \int d\vec\rho d\vec\rho' \vec M(\vec\rho) \cdot \mat G(\vec\rho,\vec\rho') \cdot \vec M(\vec\rho') \,,
  \end{split}
\end{equation}  
where $\lambda_{ex}$ is the exchange length, $R$ is the disc radius and $\mat G$ is the magnetostatic Green's function. Substituting definition of the canonical variable and its eigenmode expansion into Eq.~\eqref{e:H} and making straightforward algebra the 3-wave part of the transformed Hamiltonian is derived in the form  $\H^{(3)} = \sum_{0,1,2} (V_{12,0} a_1a_2a_0^* + \text{c.c.})\Delta(m_1+m_2-m_0)$, where we use short notations $a_1 = a_{n_1 m_1}$. The term $V_{12,0}a_1a_2a_0^*$ describes the splitting of mode $a_0$ into $a_1$ and $a_2$. The delta function represents the selection rule for the azimuthal mode numbers $m_1+m_2 = m_0$, which is a consequence of the momentum conservation law. Calculations show, that in our case the exchange interaction does not contribute to the 3-magnon coefficient $V_{12,0}$, since $V_{12,0}^{(ex)} \sim m_0 = 0$. Using the expression for Green's function in polar coordinates \cite{Guslienko_2000}, one finds the dipolar contribution in the case $m_0=0$, $m_1=-m_2=m$ equal to
  \begin{equation}\label{e:Vmm0}
  \begin{gathered}
   V_{12,0} = \frac{im\omega_M}{2\sqrt{2}R^2} \iiint \left(J_{m+1}(k\rho) - J_{m-1}(k\rho)\right) J_m(k\rho')\times \\
    f(kL) g_0(\rho') \left[g_1(\rho) g_2(\rho') - g_2(\rho) g_1(\rho')  \right] \rho d\rho d\rho'dk  \,,
   \end{gathered}
  \end{equation}
where $f(x) = 1-(1-e^{-|x|})/|x|$, $L$ is the disc thickness and $J_m$ is the Bessel function. The last term in brackets in Eq.~(\ref{e:Vmm0}) is nonzero only if $g_{n_1,m}(\rho) \neq g_{n_2,-m}(\rho)$. Radial profiles of modes having the same radial number $n$ and opposite azimuthal numbers are the same $g_{n,m}(\rho) = g_{n,-m}(\rho)$,  (except for the case $m = \pm 1$, when the hybridisation with the gyrotropic mode appears \cite{Guslienko2008}, but this difference is small being proportional to $R_\mathrm{core}/R$). Thus, 3-magnon interaction efficiency is nonzero only if $n_1 \neq n_2$. Consequently, the selection rules for splitting of $(n_0,0)$ mode into $(n_1, m_1)$ and $(n_2, m_2)$ are: $m_1=-m_2$ (consequence of momentum conservation law) and $n_1 \neq n_2$ (result of the symmetry of the dipolar interaction). The spectra of spin-wave excitations and the corresponding profiles (Fig.~\ref{fig1}b,c) were calculated as numerical solution of linearised Landau-Lifshitz equation using the projection method \cite{Buess2005}.

{\bf Micromagnetics.} Micromagnetic simulations are carried out using a custom version of MuMax3 \cite{MuMax} which uses a finite difference approach to solve the Landau-Lifshitz-Gilbert equation of motion on a rectangular grid. As a midway between accuracy and computational performance, the discs of 5.1\,$\mu$m diameter and 50\,nm thickness are modelled with a cell size of 10\,nm x 10\,nm x 5\,nm using the material parameters listed below. To determine the spatial mode amplitudes of the satellites, a harmonic out-of-plane magnetic field is applied at $f_0$ for a total duration of 100\,ns. The time-dependent magnetisation is sampled at a rate of 10\,ps and subsequently
Fourier-transformed in the time domain at each cell. From the resulting frequency spectrum, the peak positions of the satellites $f_1$ and $f_2$ are determined and the respective mode amplitudes are calculated via backwards Fourier transform at these frequencies.

Material parameters used for analytical calculations as well as for micromagnetic simulations:
thickness 50\,nm, diameter 5.1\,$\mu$m, saturation magnetisation $M_{s}=810$\,kA/m, $\gamma= 1.86\times 10^{11}$\,rad/(s T), exchange constant $A = 1.3\times 10^{-11}$\,J/m.







\section*{acknowledgments}
The authors acknowledge fruitful discussions with S.V. Kusminskiy.
Financial support by the Deutsche Forschungsgemeinschaft is gratefully acknowledged within program SCHU2922/1-1.  K.S. acknowledges funding within the Helmholtz Postdoc Programme. Samples were fabricated at the Nanofabrication Facilities (NanoFaRo) at the Institute of Ion Beam Physics and Materials Research at HZDR. We thank B. Scheumann for film deposition and L. Bischoff for the thickness measurement. 

\section*{Author contributions}
K.S. and H.S. conceived and designed the experiments. F.W. and K.W. fabricated the samples. R.V., V.T., and A.N.S. did analytical calculations. L.K. and A.K. performed micromagnetic simulations. A.A.A. performed the COMSOL simulations. K.S., F.W., K.W., T.H., L.K., and T.H. performed BLS experiments and analysed the data. K.S. and H.S. wrote the manuscript. All authors discussed the results and commented on the manuscript.

\section*{Additional information}
The authors declare no competing financial interests. Reprints and permission information is available online at http://www.nature.com/reprints. Correspondence and requests for materials should be addressed to K.S.
\end{document}